\documentclass[12pt,twoside]{article}
\usepackage{fleqn,espcrc1}
\usepackage{graphicx}
\usepackage{epsfig}
\usepackage[figuresright]{rotating}

\newcommand{\hethree}{\mbox{$^{3}$He}}

\newcommand{\hefour}{\mbox{$^{4}$He}}

\newcommand{\AmS}{{\protect\the\textfont2
  A\kern-.1667em\lower.5ex\hbox{M}\kern-.125emS}}

\title{The {\it hep} reaction and the solar neutrino problem}

\author{L.E. Marcucci\address{Department of Physics, 
	Old Dominion University, Norfolk, Virginia 23529} 
        \thanks{present address: 
	Department of Physics, University of Pisa, 
	Via Buonarroti 2, I-56100 Pisa, Italy.}}
       
\begin{document}

\maketitle

\begin{abstract}
The results of a new calculation of the 
astrophysical $S$-factor for the proton weak capture on $\hethree$ 
are here reviewed. The methods used to obtain very accurate 
initial and final state wave functions and to 
construct the nuclear weak current operator are described. 
Finally the implications of these results for the Super-Kamiokande 
solar neutrino data are discussed.
\end{abstract}

\section{INTRODUCTION}
\label{sec:intro}

In the present talk, I will report about a recent study of the 
process $\hethree(p,e^+ \nu_e)\hefour$, 
also known as $hep$ reaction~\cite{Mar00}.
This process has recently received considerable attention~\cite{Mar00,BK98}, 
triggered by the results presented by the Super-Kamiokande 
(SK) collaboration of the energy spectrum of electrons 
recoiling from scattering with solar neutrinos. In fact, while 
over most of the spectrum, a constant suppression 
of about 0.5 is observed relative to the Standard Solar Model (SSM) 
predictions~\cite{BBP98}, above 12.5 MeV there is an apparent excess of 
events. Accordingly with the SSM, the $hep$ reaction is 
the only source of solar neutrinos with energy higher than 14 MeV (their 
end-point is about 19 MeV). This fact has led to questions 
about the reliability of the calculations of the $hep$ 
reaction cross section, 
upon which the SSM bases its currently accepted value for the 
astrophysical $S$-factor, 2.3 $\times 10^{-20}$ keV b~\cite{Sch92}. 
In particular,  
the SK collaboration~\cite{Suz00} has shown that a large enhancement, 
by a factor of about 17, of the $hep$ $S$-factor would essentially 
fit the observed excess of recoiling electrons.

The theoretical description of the $hep$ process, 
as already known since long time,  constitutes a challenging 
problem from the standpoint of nuclear few-body theory. 
In fact, as discussed in detail in Ref.~\cite{Mar00}, 
the $hep$ reaction is extremely sensitive to: 
(i) small components in the wave functions, in particular the D-state 
admixtures generated by the tensor interactions; 
(ii) relativistic corrections and many-body terms in the 
weak transition operator; 
(iii) P-wave capture contributions.  

The outline of the talk is as follow: I will first 
briefly review the main steps of the calculation, in particular 
discussing the method used to describe the initial and final state 
wave functions and the model of the weak transition operators. 
Then I will present the $S$-factor results and 
discuss their implication for the SK solar neutrino spectrum.
Some concluding remarks are given in Section~\ref{sec:concl}.

\section{REVIEW OF THE CALCULATION}
\label{sec:cal}
This Section is divided in three parts: in Subsection~\ref{subsec:xs}, 
I first summarize all the relevant formulas for the $hep$ 
astrophysical $S$-factor and the cross section; a detailed description 
can be found in Ref.~\cite{Mar00}.  
In Subsection~\ref{subsec:chh}, I then discuss the 
correlated-hyperspherical-harmonics method used to describe the 
initial and final state wave functions; 
finally in Subsection~\ref{subsec:cc}, 
I present the model for the weak transition operator.

\subsection{The $hep$ cross section and astrophysical $S$-factor}
\label{subsec:xs}

The astrophysical $S$-factor at center-of-mass (c.m.) energy $E$ 
is defined as
\begin{equation}
S(E) = E\, \sigma(E)\,
{\rm e}^{2\, \pi \, \eta} \ ,
\label{eq:sfc}
\end{equation}
where $\sigma(E)$ is the $hep$ cross section and $\eta$ is given by
$\eta=2\alpha/v_{\rm rel}$, 
$\alpha$ being the fine structure constant and $v_{\rm rel}$ the 
$p\,\hethree$ relative velocity. 
The cross section $\sigma(E)$ is written as:
\begin{equation}
\sigma(E)=\int 2\pi \,\, \delta\!\left (\Delta m  + E -
\frac{q^{2}}{2 m_{4}} - E_e
- E_\nu\right )\frac{1}{v_{\rm rel}} \frac{1}{4}
\sum_{s_e s_\nu}\sum_{s_1 s_3}
|\langle f\,|\,H_{W}\,|\,i\rangle|^{2}
\frac{d{\bf{p}}_{e}}{(2\pi)^3} \frac{d{\bf{p}}_{\nu}}{(2\pi)^3} \ ,
\label{eq:xsc1}
\end{equation}
where $\Delta m = m + m_3 - m_4 $ = 19.29 MeV
($m$, $m_3$, and $m_4$ are the proton, $\hethree$, and $\hefour$ rest masses,
respectively), and the transition amplitude is given by
\begin{equation}
\langle f|H_{W}|i\rangle=
\frac{G_{V}}{\sqrt{2}}\,l^{\sigma} \langle -{\bf{q}}; ^{4}\!{\rm{He}}|
j_{\sigma}^{\dag}({\bf{q}})|{\bf{p}}; p\,^{3}{\rm{He}}\rangle \ .
\label{eq:tra}
\end{equation}
Here $G_V$ is the Fermi constant, ${\bf{q}}={\bf{p}}_{e}+{\bf{p}}_{\nu}$,
$|{\bf{p}}; p\,^{3}{\rm{He}}\rangle$ and
$|-\!{\bf{q}}; ^{4}\!{\rm{He}} \rangle$ represent, respectively,
the $p\,\hethree$ scattering state with relative momentum
${\bf{p}}$ and $\hefour$ bound state recoiling with momentum $-{\bf{q}}$,
$l_{\sigma}$ is the leptonic weak current,
$l_{\sigma} = \overline{u}_{\nu}\gamma_{\sigma}(1-\gamma_5)v_{e}$ (the
lepton spinors are normalized as 
$v_{e}^{\dag}v_{e}={u}_{\nu}^{\dag}{u}_{\nu}=1$),
and $j^\sigma({\bf q})$ is the nuclear weak current, $j^\sigma({\bf q})
=( \rho({\bf q}), {\bf j}({\bf q}))$.  The dependence of the amplitude
upon the spin projections of the leptons, proton and $^3$He has been
omitted for ease
of presentation.  

The c.m.\ energies of interest involved in the 
$p\,\hethree$ weak capture reaction, are of the order of 10 keV: 
the energy at which the reaction is most probable to occur, known as 
the Gamow-peak energy, is in fact 10.7 keV. 
Therefore, it is convenient to expand the $p\,\hethree$ scattering state
into partial waves, and perform a multipole decomposition of the nuclear
weak charge, $\rho({\bf q})$, and current, ${\bf j}({\bf q})$, operators.
Standard manipulations lead to~\cite{Mar00}
\begin{equation}
\frac{1}{4}\sum_{s_e s_\nu}\sum_{s_1 s_3}
|\langle f\,|\,H_{W}\,|\,i\rangle|^{2} = 
(2 \pi)^2\> G_V^2\> L_{\sigma \tau} \> N^{\sigma\tau} ,
\end{equation}
where the lepton tensor $L^{\sigma \tau}$ is written in terms of
electron and neutrino four-velocities, while the nuclear tensor 
$N^{\sigma\tau}$ is 
given in terms of the reduced matrix
elements (RMEs) of the Coulomb $(C_{\ell \ell_z})$,
longitudinal $(L_{\ell \ell_z})$,
transverse electric $(E_{\ell \ell_z})$,
and transverse magnetic $(M_{\ell \ell_z})$
multipole operators between the initial $p\,\hethree$ state
with orbital angular momentum $L$, channel spin $S$ ($S$=0,1),
and total angular momentum $J$, and final $\hefour$ state.
The present study includes S- and P-wave capture channels,
i.e. the $^1$S$_0$, $^3$S$_1$, $^3$P$_0$, $^1$P$_1$, $^3$P$_1$, and $^3$P$_2$
states, and retains all contributing
multipoles connecting these states to the $J^\pi$=0$^+$ ground
state of $\hefour$.

\subsection{The initial and final state wave functions}
\label{subsec:chh}

The correlated-hyperspherical-harmonics (CHH) method, developed 
for the four-body problem in
Refs.~\cite{VKR95,VRK98}, has been used to calculated the 
bound- and scattering-state wave functions. I first 
describe the method for the $\hefour$ wave function. 

\subsubsection{The $\hefour$ wave function}
\label{subsubsec:he4}

In the study of the four-nucleon systems, there are 
two sets of Jacobi coordinates, $\{ {\bf{x}}_A, {\bf{y}}_A, {\bf{z}}_A\}$ 
and $\{ {\bf{x}}_B, {\bf{y}}_B, {\bf{z}}_B\}$,  
corresponding to the partitions 1+3 and 2+2, respectively 
(note that by definition ${\bf{x}}_A = {\bf{x}}_B$). Their explicit 
expressions can be found in Refs.~\cite{Mar00,VKR95}. 
In the CHH method, the magnitudes of the Jacobi variables 
are replaced by the 
so-called hyperspherical coordinates, which in the 
four-body case are given by: 
\begin{equation}
\rho=\sqrt{ x^2_{A}+ y^2_{A}+z^2_{A}}
    =\sqrt{ x^2_{B}+ y^2_{B}+z^2_{B}}\ , 
\label{eq:hypr} 
\end{equation}
\begin{eqnarray}
       \cos \phi_{3}&=&  x_{A}/\rho =  x_{B}/\rho\ , \\
 \noalign{\medskip}
      \cos \phi^A_{2}&=&  y_{A}/(\rho \sin\phi_{3})  \ , \\
 \noalign{\medskip}
      \cos \phi^B_{2}&=& y_{B}/(\rho \sin\phi_{3} ) \ .
       \label{eq:hypzx}
\end{eqnarray}
The $\hefour$ wave function can be now expanded as:
\begin{equation}
\Psi=\sum_{n}\frac{z_{n}(\rho)}{\rho^4}
Z_{n}(\rho,\Omega) \ , \label{eq:psi}
\end{equation}
where $z_{n}(\rho)$ are hyper-radial functions, yet to be determined, and 
$Z_{n}(\rho,\Omega)$ are known functions, which contain all the 
spin, isospin, angle and hyper-angle dependence and a Jastrow correlation 
factor. This factor accounts for the strong state-dependent correlations 
induced by the nucleon-nucleon interaction and  
improves the behaviour of the wave function at small interparticle 
distances, thus accelerating the convergence of the calculated quantities 
with respect to the number of required basis functions. 
The hyper-radial functions $z_n(\rho)$ and the bound 
state energy $E$ are then obtained  
applying the Rayleigh-Ritz variational principle, 
$\langle\delta_z\Psi | H-E |  \Psi \rangle =0$.
The nuclear Hamiltonian $H$ 
consists here of the Argonne $v_{18}$ two-nucleon~\cite{WSS95} and Urbana-IX
three-nucleon~\cite{Pud95} interactions. 
To make contact with earlier
studies~\cite{Sch92,Car91}, however, and to
have some estimate of the model dependence
of the results, the older Argonne $v_{14}$ two-nucleon~\cite{WSA84}
and Urbana-VIII three-nucleon~\cite{Wir91}
interaction models have also been used. 
Both these Hamiltonians, the AV18/UIX and AV14/UVIII, reproduce
the experimental binding energies and charge
radii of the trinucleons and $\hefour$ in exact Green's function
Monte Carlo (GFMC) calculations~\cite{Pud97,Car90a}. 
The results of the $\hefour$ binding energy calculated 
with the CHH method are given in Table~\ref{tb:be} and compared 
with the GFMC values.  
Depending on the Hamiltonian model, 
the CHH results~\cite{VKR95,Viv99b} are within 1--2 \%,
of those obtained with the GFMC method.  

\begin{table}[htb]
\caption{Binding energies in MeV of $^{4}$He calculated with the CHH method
using the AV18 and AV18/UIX, and the older AV14 and AV14/UVIII, 
Hamiltonian models.  Also listed are the corresponding 
\lq\lq exact\rq\rq GFMC results~\protect\cite{Pud97,Car90a} 
and the experimental value.}
\label{tb:be}
\newcommand{\m}{\hphantom{$-$}}
\newcommand{\cc}[1]{\multicolumn{1}{c}{#1}}
\renewcommand{\tabcolsep}{2pc} 
\renewcommand{\arraystretch}{1.2} 
\begin{tabular}{@{}lcc}
\hline
    Model    &  CHH & GFMC  \\
\hline
AV18     &  24.01  & 24.1(1)  \\
AV18/UIX   & 27.89 & 28.3(1)  \\
AV14     &  23.98  & 24.2(2)  \\
AV14/UVIII & 27.50 & 28.3(2)  \\
\hline
EXP&\multicolumn{2}{c}{28.3} \\
\hline
\end{tabular}
\end{table}

\subsubsection{The $p\,\hethree$ wave function}
\label{subsubsec:phe3}

The $p\,\hethree$ cluster wave function $\Psi_{1+3}^{LSJJ_z}$, having
incoming orbital angular momentum $L$ and channel spin $S$ ($S=0, 1$)
coupled to total angular $JJ_z$, is expressed as 
\begin{equation}
\Psi_{1+3}^{LSJJ_z}=\Psi_C^{JJ_z}+\Psi_A^{LSJJ_z} \ ,
\label{eq:psica}
\end{equation}
where the term $\Psi_C$ vanishes in the
limit of large intercluster separations, and
hence describes the system in the region where the particles are close
to each other and their mutual interactions are strong.
The term $\Psi_A^{LSJJ_z}$ describes the system
in the asymptotic region, where proton and 
$\hethree$ interact only via the Coulomb interaction. 
It contains the dependence on the $R$-matrix elements, 
which determine phase shifts and (for coupled
channels) mixing angles, and it is written in terms of the  
$\hethree$ wave function, which is obtained using the same CHH method 
as discussed above, but for a three-body systems~\cite{Kie94,Mar98}.

The \lq\lq core\rq\rq wave function $\Psi_C$ is expanded in the
same CHH basis as the bound-state wave function, and both
the $R$-matrix elements and the 
functions $z_{n}(\rho)$ occurring in the expansion of
$\Psi_C$ are determined applying the Kohn variational 
principle~\cite{Mar00,VKR95}.  

The $\hethree$ binding energy and the 
$p\,\hethree$ singlet and triplet scattering lengths predicted
by the Hamiltonian models considered in the present work
are listed in Table~\ref{tb:scl}, and are found in good agreement
with available experimental values, although these are rather
poorly known.  The experimental scattering lengths have been
obtained, in fact, from effective range parametrizations
of data taken above $1$ MeV, and therefore might have 
large systematic uncertainties. 

\begin{table}[htb]
\caption{Binding energies, $B_3$, of $\hethree$, and $p\,\hethree$ singlet
and triplet S-wave scattering lengths, $a_{\rm s}$ and
$a_{\rm t}$, calculated with the CHH method
using the AV18 and AV18/UIX, and the older AV14 and AV14/UVIII, 
Hamiltonian models.  The corresponding experimental values are also listed.}
\newcommand{\m}{\hphantom{$-$}}
\newcommand{\cc}[1]{\multicolumn{1}{c}{#1}}
\renewcommand{\tabcolsep}{2pc} 
\renewcommand{\arraystretch}{1.2} 
\begin{tabular}{@{}lccc}
\hline
Model      & $B_3$(MeV)  & $a_{\rm s}$(fm) & $a_{\rm t}$(fm) \\
\hline
AV14       & 7.03 &         &        \\
AV18       & 6.93 &  12.9   & 10.0   \\
AV14/UVIII & 7.73 &         & 9.24   \\
AV18/UIX   & 7.74 &  11.5   & 9.13   \\
\hline
EXP        & 7.72 &  10.8$\pm$2.6~\protect\cite{AK93} 
                  & 8.1$\pm$0.5~\protect\cite{AK93} \\
           &      &         & 10.2$\pm$1.5~\protect\cite{TEG83} \\
\hline
\end{tabular}
\label{tb:scl}
\end{table}

\subsection{The nuclear weak current}
\label{subsec:cc}

The nuclear weak current 
$j^\sigma({\bf{q}})=(\rho({\bf{q}}),{\bf{j}}({\bf{q}}))$ 
has vector $(V)$ and axial-vector $(A)$ parts, with
corresponding one- and many-body components.  All the one-body
terms can be obtained in a standard way from a non-relativistic
reduction of the covariant single-nucleon vector and axial-vector
currents, including terms proportional to $1/m^2$.
The two-body components of the weak vector current ${\bf{j}}({\bf{q}};V)$ 
are constructed from the isovector two-body 
electromagnetic currents in accordance with the conserved-vector-current
(CVC) hypothesis, and consist~\cite{Mar00} of \lq\lq model-independent\rq\rq
(MI) and \lq\lq model-dependent\rq\rq (MD) terms.  The MI
terms are obtained from the nucleon-nucleon interaction, and by
construction satisfy current conservation with it.
The leading MI two-body contribution is given by 
the \lq\lq$\pi$-like\rq\rq operator, obtained from the isospin-dependent
spin-spin and tensor nucleon-nucleon interactions.
The latter also generate an isovector \lq\lq$\rho$-like\rq\rq current,
while additional isovector two-body currents arise 
from the isospin-independent
and isospin-dependent central and momentum-dependent interactions.  These
currents are short-ranged, and numerically far less important
than the $\pi$-like current.  With the exception of the $\rho$-like current,
they have been neglected in the present work.  The MD 
currents are purely transverse, and therefore cannot be directly
linked to the underlying two-nucleon interaction.  The present
calculation includes the currents associated with excitation
of $\Delta$ isobars which, however, are found to give a rather small
contribution in weak-vector transitions, as compared to that due to
the $\pi$-like current.  

The many-body weak charge operators can also be obtained from their 
electromagnetic correspondents applying the CVC hypothesis. However, 
while the main parts of the two-body electromagnetic or weak vector
current are linked to the form of the nucleon-nucleon
interaction through the continuity equation, the most important two-body
electromagnetic or weak vector charge operators are
model dependent, and should be viewed as relativistic
corrections. 
The model commonly used~\cite{Sch90} for the electromagnetic 
many-body charge operators includes the $\pi$-, $\rho$-, and
$\omega$-meson exchange terms with both isoscalar and isovector
components, as well as the (isoscalar) $\rho \pi \gamma$ and (isovector)
$\omega \pi \gamma$ charge transition couplings (in addition to the
single-nucleon Darwin-Foldy and spin-orbit relativistic corrections).
The $\pi$- and $\rho$-meson exchange charge operators are constructed from the
isospin-dependent spin-spin and tensor interactions, using the same
prescription adopted for the corresponding current operators~\cite{Sch90}.
At moderate values of momentum transfer ($q \!< \! 5$ fm$^{-1}$), the
contribution due to the \lq\lq $\pi$-like\rq\rq 
exchange charge operator has been found to
be typically an order of magnitude larger
than that of any of the remaining two-body mechanisms
and one-body relativistic corrections~\cite{Mar98}.  
In the present study therefore we
retain, in addition to the one-body operator, 
only the \lq\lq $\pi$-like\rq\rq and 
\lq\lq $\rho$-like\rq\rq weak vector charge operators.

The axial charge operator $\rho({\bf{q}}; A)$ includes, in
addition to the one-body component, 
the long-range pion-exchange term~\cite{Kub78}, required by low-energy
theorems and the partially-conserved-axial-current relation,
as well as the (expected) leading short-range terms constructed 
from the central
and spin-orbit components of the nucleon-nucleon interaction, following
a prescription due to Kirchbach {\it et al.}~\cite{Kir92}.
The $\Delta$-excitation terms have also been included, but they have been 
found to be unimportant~\cite{Mar00}. 

In contrast to the electromagnetic case, the axial current operator 
${\bf{j}}({\bf{q}}; A)$ is not
conserved.  Thus, 
its two-body components cannot be linked to the nucleon-nucleon
interaction and, in this sense, should be viewed as model dependent.
In the model presented here, the
two-body axial current operators due to $\pi$- and $\rho$-meson
exchanges,
and the $\rho\pi$-transition mechanism have been included.
The leading many-body terms in the
axial current are however due to $\Delta$-isobar
excitation. They have been treated non-perturbatively in the 
transition-correlation-operator (TCO) scheme, originally
developed in Ref.~\cite{Sch92}
and further extended in Ref.~\cite{Mar98}.  In the
TCO scheme--essentially, a scaled-down approach to
a full $N$$+$$\Delta$ coupled-channel treatment--the
$\Delta$ degrees of freedom are explicitly
included in the nuclear wave functions.  

The largest model dependence is in the weak axial current.  To minimize it,
the poorly known $N$$\Delta$ transition
axial coupling constant $g_A^*$ has been
adjusted to reproduce the experimental value
of the Gamow-Teller matrix element in tritium
$\beta$-decay~\cite{Mar00,Sch98}.  While this
procedure is model dependent, its actual model dependence is in fact
very weak, as has been shown in Refs.~\cite{Mar00,Sch98}. 

\section{RESULTS}
\label{sec:res}
I present here the results for the $hep$ astrophysical 
$S$-factor, and their 
implications to the SK solar neutrino spectrum. 

\subsection{Results for the $S$-factor}
\label{subsec:res_sfc}
The results for the astrophysical $S$-factor, calculated using 
CHH wave functions with the AV18/UIX Hamiltonian model, 
at three different c.m.\ energies, are given in Table~\ref{tb:sfact}.  By 
inspection of the table, it can be noted that: (i) the energy dependence is
rather weak: the value at $10$ keV is only about 4 \% larger
than that at $0$ keV; (ii) the P-wave capture states are found to
be important, contributing about 40 \% of the calculated
$S$-factor.  However, the contributions from D-wave channels
are expected to be very small.  It has been explicitly verified 
that they are indeed small in $^3$D$_1$ capture. 
(iii) The many-body axial currents play a crucial
role in the (dominant) $^3$S$_1$ capture, where they reduce
the $S$-factor by more than a factor of four. 

\begin{table}[bht]
\caption{The $hep$ $S$-factor, in units of $10^{-20}$ keV~b, calculated
with CHH wave functions corresponding to the AV18/UIX Hamiltonian model,
at $p\,\hethree$ c.m.\ energies $E$=0, 5, and 10 keV.  The rows
labelled \lq\lq one-body\rq\rq and \lq\lq full\rq\rq list the
contributions obtained by retaining the one-body only and both
one- and many-body terms in the nuclear weak current.  The contributions due 
the $^3$S$_1$ channel only and all S- and P-wave channels are
listed separately.}
\begin{tabular}{ccccccc}
\hline
& \multicolumn{2}{c} {$E$=$0$ keV} &
  \multicolumn{2}{c} {$E$=$5$ keV} &
  \multicolumn{2}{c} {$E$=$10$ keV} \\
& $^3$S$_1$ & S+P & $^3$S$_1$ & S+P & $^3$S$_1$ & S+P\\
\hline
one-body  &26.4  & 29.0 & 25.9 & 28.7 & 26.2 & 29.3 \\
full      &6.38  & 9.64 & 6.20 & 9.70 & 6.36 & 10.1 \\
\hline
\end{tabular}
\label{tb:sfact}
\end{table}

The different contributions from the S- and P-wave capture channels to 
the zero energy $S$-factor are given in Table~\ref{tb:sfc_contr}. The 
results obtained using the two-nucleon AV18 and the older two- 
and three-nucleon AV14/UVIII interaction models are also 
listed. Note 
that the sum of the channel contributions is a few \% smaller than the 
total result reported at the bottom of the table, due to the presence of  
interference terms among multipole operators connecting 
different capture channels~\cite{Mar00}. 

The dominant contribution to the $S$-factor is 
obtained from the $^3$S$_1$ capture channel. Among the P-wave 
capture channels, the $^3$P$_0$ does not give the largest contribution, 
as instead expected in previous studies~\cite{BK98}, although 
this is the only contribution surviving in the limit $q$=0. 
\begin{table}[hbt]
\caption{Contributions of the S- and P-wave capture channels to 
the $hep$ $S$-factor at zero $p\,\hethree$ c.m. energy in 10$^{-20}$ keV~b. 
The results correspond to the AV18/UIX, AV18 and   
AV14/UVIII Hamiltonian models.}
\begin{tabular}{cccc}
\hline
& AV18/UIX & AV18 & AV14/UVIII \\
\hline
$^{1}$S$_{0}$ & 0.02 & 0.01 & 0.01\\
$^{3}$S$_{1}$ & 6.38 & 7.69 & 6.60\\
$^{3}$P$_{0}$ & 0.82 & 0.89 & 0.79\\
$^{1}$P$_{1}$ & 1.00 & 1.14 & 1.05\\
$^{3}$P$_{1}$ & 0.30 & 0.52 & 0.38\\
$^{3}$P$_{2}$ & 0.97 & 1.78 & 1.24\\
\hline
TOTAL         & 9.64 & 12.1 & 10.1 \\
\hline
\end{tabular}
\label{tb:sfc_contr}
\end{table}

By comparing the AV18 and AV18/UIX results, it can be concluded that  
inclusion of the three-nucleon interaction reduces the 
total $S$-factor by about 20 \%.  This decrease
is mostly in the $^3$S$_1$ contribution, and can be traced back
to a corresponding reduction in the magnitude of the one-body
axial current matrix elements.  The latter are sensitive to the
triplet scattering length, for which the AV18 and AV18/UIX
models predict, respectively, 10.0 fm and 9.13 fm (see Table~\ref{tb:scl}).
This 20 \% difference in the total $S$-factor values for AV18 and 
AV18/UIX emphasizes the need for performing the 
calculation using a Hamiltonian model that reproduces 
the binding energies and low-energy scattering parameters 
for the three- and four-nucleon systems. This is true for the AV18/UIX 
model, but not for the AV18 model.

The different contributions to the astrophysical $S$-factor when the older 
AV14/UVIII potential model is used are given in the last column of 
Table~\ref{tb:sfc_contr}. By comparing these results with the ones 
obtained with the AV18/UIX, it can be observed 
that both the S- and P-wave contributions are not 
significantly changed; in particular, 
the $^3$S$_1$ capture $S$-factor values differ for only 
about 3 \%. It is important to emphasize that this 
is due to the procedure of 
constraining the model dependent two-body axial currents 
by fitting the Gamow-Teller matrix element of 
tritium $\beta$-decay, as discussed at the end of the previous 
Section. Note that the AV14/UVIII Hamiltonian 
also reproduces the low-energy properties for the three- and 
four-nucleon systems. 

\subsection{Implications for the Super-Kamiokande solar neutrino spectrum}
\label{subsec:res_sk}
\begin{figure}[bht]
\caption{Electron energy spectrum for the ratio 
between the Super-Kamiokande 825-days data and the expectation 
based on unoscillated $^8$B neutrinos~\protect\cite{BBP98}. The data 
were extracted graphically from Fig. 8 of Ref.~\protect\cite{Suz00}. 
The 5 curves correspond respectively to no $hep$ contribution (dotted line), 
and an enhancement $\alpha$ of 2.2 (solid line), 4.4 (long-dashed line), 
10 (dashed line) and 20 (dot-dashed line).}
\epsfig{file=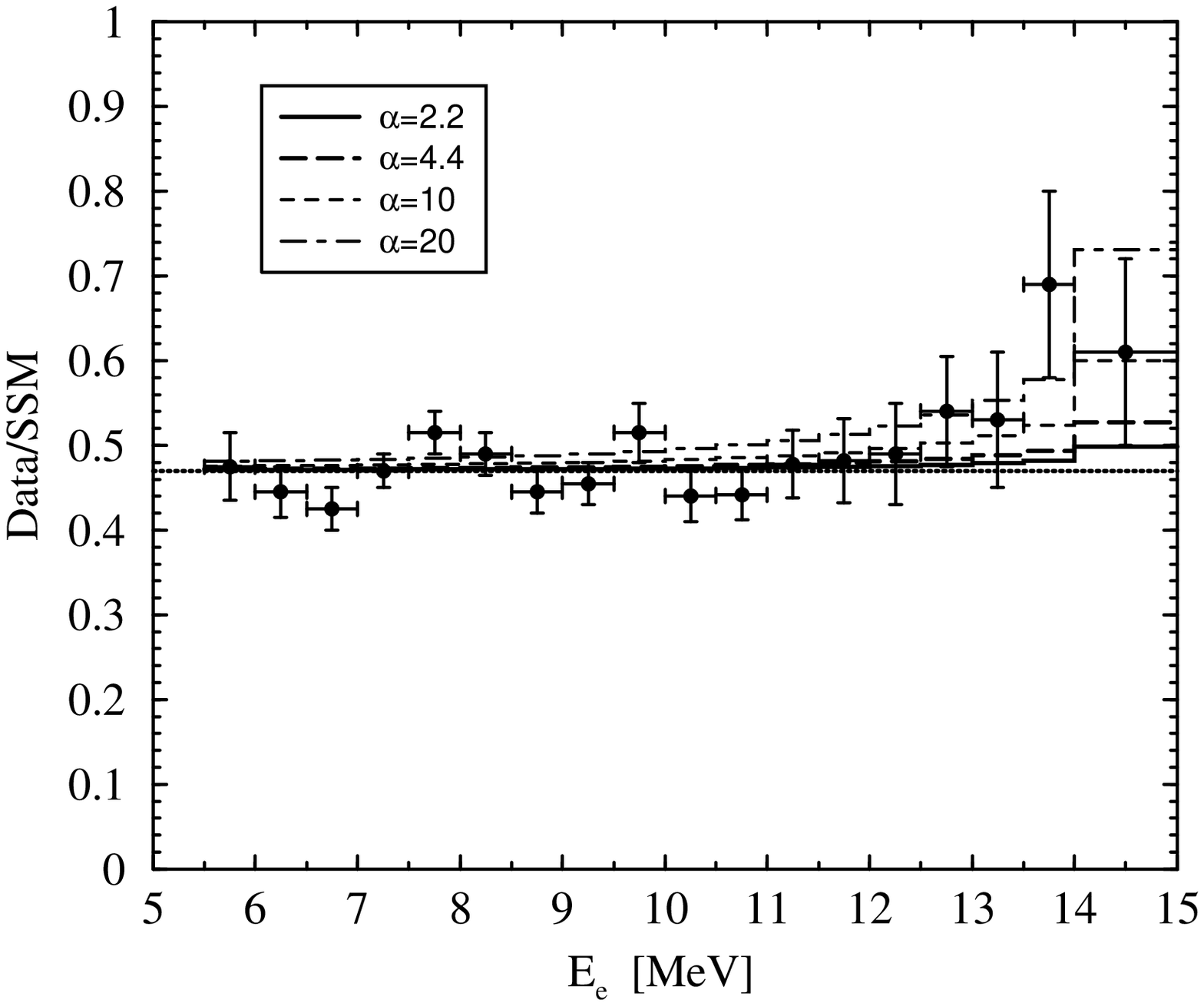,height=12cm,width=14cm}
\label{fig:ratio}
\end{figure}

The Super-Kamiokande (SK) experiment detects 
solar neutrinos by neutrino-electron
scattering.  It is sensitive, according to the SSM~\cite{BBP98}, 
to the very energetic neutrinos from the $^8$B weak decay 
($^8{\rm B}\rightarrow \hefour +\hefour +e^+ +\nu_e$) and from the 
$hep$ reaction. 
The SK results are presented as ratio of the measured  to the SSM predicted 
events when no neutrino oscillations are included, as function of the 
recoil electron energy. 
Over most of the spectrum, this ratio is constant at 
$\simeq 0.5$~\cite{Suz00}. At
the highest energies, however, there is an excess of events 
relative to the $0.5 \times$SSM prediction. 
This is seen in Fig.~\ref{fig:ratio} where the SK
results from 825 days of data acquisition~\cite{Suz00}
are shown by the points (the error bars denote the
combined statistical and systematic error); the dotted line is the 
$0.5 \times$SSM prediction.   

To study the effects of the new value for the $S$-factor presented here, 
10.1 $\times 10^{-20}$ keV~b (see Table~\ref{tb:sfact}) to the SK spectrum, 
it is useful to introduce 
the ratio $\alpha$ of the $hep$ flux to its SSM value, defined as 
$\alpha \equiv S_{\rm new}/S_{\rm SSM} \times P_{\rm osc}$, 
where $P_{\rm osc}$ is the observed suppression factor due to 
neutrino oscillations. Therefore, 
if $hep$ neutrino oscillations are ignored, then
$\alpha = (10.1\times 10^{-20}{\rm\ keV~b})/(2.3\times 10^{-20}{\rm\
keV~b}) = 4.4$, while if the $hep$ neutrinos are suppressed
by $\simeq 0.5$, then $\alpha = 2.2$. 
The long-dashed and solid lines in Fig.~\ref{fig:ratio} 
indicate the effect of these two different values of $\alpha$ 
on the ratio of the electron spectrum with
both $^8$B and $hep$ to that with only $^8$B (the SSM).  
Two other arbitrary values
of $\alpha$ (10 and 20) are shown for comparison. 

\section{CONCLUSIONS}
\label{sec:concl}
In this talk, I have reported about a recent new calculation 
of the astrophysical $S$-factor for the $hep$ reaction. 
The chief conclusion of this calculation is that the  
best estimate for the $S$-factor at 
10 keV, close to the Gamow-peak energy, is 10.1 $\times 10^{-20}$ keV~b. 
This value is $\simeq$ 4.5 times larger than the value adopted in SSM, 
based on Ref.~\cite{Sch92}, of 2.3 $\times 10^{-20}$ keV~b. 
It is therefore important to point out the differences between the present
and the previous study of Ref.~\cite{Sch92}: 
(i) all P-wave contributions are included; 
(ii) the CHH method has been used to describe the initial and 
final state wave functions, corresponding to the latest generation of 
realistic interactions. The CHH method is known to be more accurate 
than the variational Monte Carlo (VMC) 
technique used in Ref.~\cite{Sch92}, and it better 
describes the small components of the wave function to which the 
one-body axial current operator is most sensitive. 
(iii) The $1/m^2$ relativistic
corrections in the one-body axial current operator are included. 
In $^3$S$_1$
capture, for example, these terms increase by 25 \% the
$L_1$ and $E_1$ matrix elements calculated with the
one-body axial current operator. 

Finally, the implications of this new estimate for the 
SK solar neutrino data have been investigated. The results 
are summarized in Fig.~\ref{fig:ratio}, from which it can be concluded that 
the enhancement of the $S$-factor reported here, although 
large, is not enough to completely resolve the discrepancies 
between the present SK results and the SSM predictions. 
However, this accurate 
calculation of the $S$-factor, and the consequent
absolute prediction for the $hep$ neutrino flux, will allow much
greater discrimination among the proposed solutions to this 
problem, based on different solar neutrino oscillation 
scenarios.

\section{ACKNOWLEDGMENTS}
I wish to thank R.\ Schiavilla, M.\ Viviani, A.\ Kievsky, 
S.\ Rosati and J.F.\ Beacom for their many important 
contributions to the work reported here. I also would like to gratefully 
acknowledge the support of the U.S. Department of Energy under 
Contract No. DE-AC05-84ER40150.

\end{document}